\documentclass[prl,amsmath,twocolumn]{revtex4-1}
\usepackage[normalem]{ulem}
\usepackage{epsfig}
\usepackage{amsfonts} 
\usepackage{nicefrac}
\usepackage{color} 
\usepackage{hyperref}
\setcitestyle{square}
\usepackage{subfigure} 
\usepackage{natbib} 
\usepackage{graphicx,amssymb} 

\newcommand{\ignore}[1]{} 

\def\eq#1{${#1}$} 
\def\EQ#1#2{\begin{equation}{#1}\label{#2}\end{equation}} 
 
\newcommand{\Tau}{\tau^*}
\newcommand{\Intd}{{\rm{d}}} 
 
\newcommand{\Hquad}{\hspace{0.5em}} 
\newcommand{\HHquad}{\hspace{0.25em}} 
\usepackage{mathrsfs}
\usepackage{commath} 
\newcommand{\OO}{{\mathcal{O}}} 
 \newcommand{\ZZ}{{\mathcal{Z}}}

\usepackage{gensymb} 

\setcitestyle{square}
\usepackage{subfigure}

\usepackage{enumitem}

\begin{document}

\title{From Non-normalizable Boltzmann-Gibbs statistics to infinite-ergodic theory}
\author{Erez Aghion\textsuperscript{a,b},  David A. Kessler\textsuperscript{a}, Eli Barkai\textsuperscript{a,b}}
\affiliation{\makebox[\textwidth][c]{a) Department of Physics, b) Institute of Nanotechnology and Advanced Materials, Bar-Ilan University, Ramat-Gan 52900, Israel} }
%
%
\begin{abstract}
{We study a particle immersed in a heat bath, in the presence of an external force which decays at least as rapidly as \eq{1/x}, for example a particle interacting with a surface through a Lennard-Jones or a logarithmic potential. As time increases, our system approaches a non-normalizable Boltzmann state. We study observables, such as the energy, which are integrable with respect to this asymptotic thermal state, calculating both time and ensemble averages. We derive  a useful canonical-like ensemble which is defined out of equilibrium, using a maximum entropy principle, where the constraints are: 
normalization, finite averaged energy and a mean-squared displacement which increases linearly with time. Our work merges infinite-ergodic theory with Boltzmann-Gibbs statistics, thus extending the scope of the latter while shedding new light on the concept of ergodicity. 
}
\end{abstract}

\maketitle{}


The equilibrium state of a particle confined in a binding potential \eq{V(x)}, coupled to a heat bath of temperature \eq{T}, is given by the Boltzmann-Gibbs distribution \eq{\lim_{{t}\rightarrow\infty}P_t(x)=\frac{1}{Z}\exp(-V(x)/k_BT)}. 
Here \eq{Z=\int_{-\infty}^\infty\exp(-V(x)/k_BT)\Intd x} is the normalizing partition function, \eq{k_B} is the Boltzmann constant. Thus while the particle's trajectory, \eq{x(t)}, is erratic 
due to the interaction with its surroundings, the long-time limit of the probability density, \eq{P_t(x)}, can be predicted. When this applies, the sample mean of an observable \eq{\mathcal{O}(x)} is given by the ensemble-average \eq{\langle \mathcal{O}(x)\rangle=\int_{-\infty}^\infty \mathcal{O}(x) P_{\infty}(x)\Intd x}. From the trajectory \eq{x(t)} of a single  particle, one can calculate the time-average \eq{\overline{\mathcal{O}[x(t)]}=(1/t)\int_0^t \mathcal{O}[x(t')]\Intd t'}. According to the ergodic hypothesis, as \eq{t\rightarrow\infty}, \eq{\overline{\mathcal{O}[x(t)]}/\langle \mathcal{O}(x)\rangle\rightarrow1}. This fundamental result allows us to estimate the statistical behavior of the single particle from the properties of the ensemble and vice-versa. 

But what can we say about the thermodynamic and ergodic properties of the system when the potential is such that the Boltzmann distribution is non-normalizable? 
There are many physically important examples of this, among them:  
a particle in a Coulomb field (the ``hydrogen atom" \cite{fermi1924wahrscheinlichkeit}), a particle in the presence of a long charged rod or polymer, which can be a model of a stretched DNA,  giving rise to a logarithmic potential (see, e.g., \cite{fogedby2007dna}), and models of glassy dynamics when the density of states grows exponentially, namely Bouchaud's trap model \cite{bouchaud1992weak} (see also \cite{martin2018diffusion,ornigottibrownian}). Non-normalizable asymptotic states 
have been previously discussed in the physics literature, e.g. in the context of L\'evy walks and laser-cooled atoms  \cite{kessler2010infinite,lutz2013beyond,rebenshtok2014infinite,aghion2017large}, weakly-chaotic maps \cite{korabel2009pesin,akimoto2010role} and nonlinear oscillators \cite{meyer2017infinite}.
In this letter we advance a new thermodynamic approach for a wide class of systems out of equilibrium, where a non-normalized state takes a central role, resembling that of the Boltzmann distribution in standard thermodynamics. We explain how this state can be realized in the laboratory and what are the insights it provides. 

 Imagine the 3-dimensional motion of a particle in a viscous fluid at room temperature, which constitutes a thermal bath, floating above a flat surface. Here, \eq{x>0} is the distance between the particle and the surface.    
The surface-single molecule potential 
is \eq{V(x)}, where \eq{V(x)\rightarrow 0}, as \eq{x\rightarrow\infty}. The motion parallel to the \eq{y-z} plane is purely diffusive. The time scale of the experiment is such that the fluid bath can be assumed infinite in all dimensions. Such a motion, which is a model for molecules ``hopping'' from the membrane of a living cell \cite{bychuk1995anomalous,chechkin2012bulk,metzler2014anomalous}, recently became measurable using novel detection techniques such as \cite{pavani2009three,campagnola2015superdiffusive,krapf2016strange,wang2017three} {(for a schematic description see the supplemental material (SM))}. Since the force field is zero at large distances from the surface, clearly here \eq{Z=\int_{0}^\infty\exp(-V(x)/k_BT)\Intd x} is divergent, regardless of the temperature. 
Can one still infer the potential field from the Boltzmann factor
$\exp( - V(x)/k_b T)$, when the equilibrium state is non-normalizable? How do we formulate  the  maximum entropy principle in this case? 
Below, we show that, for observables integrable with respect to it, both the time- and the ensemble-averages are obtained directly from the non-normalizable Boltzmann state. This is reminiscent of a generalized form of ergodicity, previously discovered in the context of deterministic maps, known as infinite-ergodic theory \cite{aaronson1997introduction,Aaronson2005ergodic,thaler2006distributional} (see also \cite{akimoto2008generalized,korabel2009pesin,meyer2017infinite}). 
In addition, our new non-normalizable approach allows us to study other thermodynamical properties of the system, such as the entropy-energy relation, and generalize the  virial theorem. Thus the non-normalized state describes a very wide range of physical observables.

\begin{figure}[t]
\centering
\includegraphics[width=1.0\linewidth]{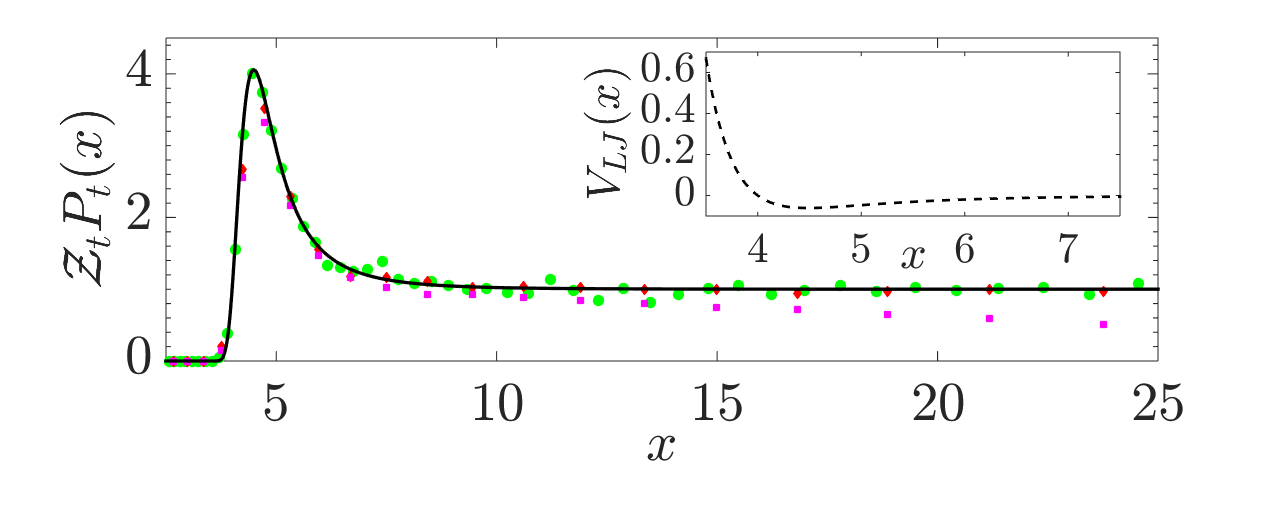}
\caption{{\footnotesize{(Color online) The 
scaled time-dependent particle density \eq{\mathcal{Z}_tP_t(x)}, where \eq{\mathcal{Z}_t=\sqrt{\pi Dt}}, obtained from simulations corresponding to the LJ potential, Eq. \eqref{Potentials}, with \eq{a,b=(2,1)}, \eq{V_0=1000} (in dimensionless units) and \eq{D=0.0436}, at times \eq{t=4700,10^5,10^6} (magenta squares, red diamonds and green circles, respectively). These results converge nicely to the non-normalizable Boltzmann state, Eq. \eqref{ID} (black line). The inset shows the  potential.  \eq{k_BT} was set to be \eq{1/4}th of the maximal potential depth.
}}}
\label{fig1}
\end{figure}

\textit{Model.} 
We model the dynamics of the diffusing particles using the unidimensional Langevin equation, in the overdamped approximation  
\EQ{\dot{x}(t)=-{V'}(x)/\gamma+\sqrt{2D}\Gamma(t).}{Langevin} 
Here~\eq{\gamma>0} is the friction constant, and according to the Einstein relation, \eq{D=k_BT/\gamma} is the diffusion coefficient. The fluctuations quantified by the second term on the right-hand side are treated as Gaussian white-noise with zero mean and \eq{\langle\Gamma(t)\Gamma(t')\rangle=\delta(t-t')}.
The spatial spreading of the diffusing particle packet is described by the Fokker-Planck equation  \cite{kubo2012statistical} \EQ{\frac{\partial P_{t}(x)}{\partial {t}}=D\left[\frac{\partial^2}{\partial {x}^2}+\frac{\partial}{\partial x} \frac{V'(x)}{k_BT}\right] P_{t}(x).}{FokkerPlanck}  
We begin by considering the large class of potential fields which are asymptotically flat at large \eq{x}. This means that, far from the surface, the force is negligibly small. As a test example, we treat in our simulations the case of a particle lightly confined by the Lennard-Jones (LJ)-type potential
\begin{equation}
V_{LJ}(x)=\begin{cases} \infty,\qquad\qquad\qquad\qquad\quad x\leq0\\ V_0(a^{12}/x^{12}-b^6/x^6),\quad\HHquad x>0 
\end{cases} 
\label{Potentials} 
\end{equation} 
where \eq{a,b,V_0>0}. This potential is plotted in the inset of Fig. \ref{fig1}.  
Potentials of this type are weak in the sense that at any finite time the system has a non-negligible population of particles that migrate to increasingly large \eq{x}, where the attractive force derived from them does not compensate for the thermal fluctuations. As a result, the density in the vicinity of the potential minimum decays in time.
   

\textit{Non-normalizable Boltzmann state.} 
We now use a scaling ansatz to solve the 
time-dependent Fokker-Planck equation, Eq. \eqref{FokkerPlanck}, for a one-sided potential \eq{V(x)}, such that \eq{V(\infty)=0}, \eq{V(0)=\infty}.  At long times and finite \eq{x}, a family of solutions to Eq. \eqref{FokkerPlanck} (up to leading order) is \eq{P_t(x)\propto t^{-\alpha}\exp(-V(x)/k_BT)}, where \eq{\alpha>0}. This is however non-normalizable, hence it cannot be the full form of the probability density. At finite times, this solution is valid only for  \eq{x/\sqrt{4Dt}\ll1}. Since \eq{V(x)} is negligibly small at large \eq{x}, \eq{P_t(x)} expands diffusively at the tails. Hence, for \eq{x\gg1}, \eq{P_t(x)\approx t^{-1/2}\exp(-x^2/(4Dt))/\sqrt{\pi D}}. 
Matching the two limits of \eq{P_t(x)},  we obtain the uniform approximation: 
  \EQ{P_t(x)\approx \frac{1}{\sqrt{\pi Dt}}e^{-V(x)/k_BT-x^2/(4D t)},}{UniformPDF} 
so \eq{\alpha=1/2} in this case. For a finite \eq{t}, the Gaussian fall-off ensures the normalizability of the probability density, although the normalization of Eq. \eqref{UniformPDF} is only approximately unity in the long-time limit. As \eq{t\rightarrow\infty}, \eq{\exp\left(-x^2/(4Dt)\right)\rightarrow1}, and the asymptotic shape of the probability density is given by the non-normalizable Boltzmann state: \EQ{\lim_{t\rightarrow\infty} \mathcal{Z}_tP_t(x)=e^{-V(x)/k_BT}, \quad\mbox{where}\quad \mathcal{Z}_t=\sqrt{\pi Dt}.}{ID} 
Importantly, this solution is independent of the shape of the (assumed narrow) initial distribution \footnote{A similar effect was shown for unstable potentials in \cite{martin2018diffusion,ornigottibrownian}}. This also holds for any asymptotically flat potential, regardless of its particular shape at finite \eq{x} 
\footnote{See the SM for technical details.}. Eq. \eqref{ID} can also be derived via eigenfunction expansion of Eq. \eqref{FokkerPlanck} \cite{dechant2011solution}. 
Eq. (5) shows that the spatial Boltzmann factor is reached in the long time limit, even though the system is not in an equilibrium state. 
Fig. \ref{fig1} shows the agreement between the asymptotic limit function of \eq{P_t(x)}, obtained from simulation results  \footnote{Simulations were performed as standard Euler-Mayurama integration of Eq.~\eqref{Langevin} 
[26].}, and  Eq. \eqref{ID},  corresponding to the LJ potential, Eq. (\ref{Potentials}). 


\textit{Ensemble-averaged observables.} Consider the ensemble-averaged observable  \eq{\langle{\mathcal{O}(x)}\rangle_t=\int_0^\infty \mathcal{O}(x)P_t(x)\Intd x}, at time \eq{t}. Using Eq. (\ref{UniformPDF}), the same argument which yielded Eq. \eqref{ID} now implies that at long times 
 \EQ{\langle \OO(x)\rangle_t\sim \frac{1}{\mathcal{Z}_t}\int_{0}^\infty \OO(x)e^{-V(x)/k_BT}\Intd x.}{EA} 
Eq. \eqref{EA} means that, similar to the case of a strongly confining field, where \eq{Z} is time-independent, the ensemble-average is obtained by integrating with respect to the non-normalized Boltzmann factor, provided that the integral exists. 
One example of such an observable is the {potential energy of the particles; \eq{E_p=V(x)}. Here, since it is zero at large \eq{x}, when applied in Eq. \eqref{EA}, \eq{V(x)} cures the non-integrability of the Boltzmann factor. 

Another physically important integrable observable is the {indicator function; \eq{\Theta(x_a<x(t')<x_b)\equiv1} when \eq{x(t')\in[x_a,x_b]}, and zero otherwise. 
Here, the ensemble-average \eq{\langle\Theta(x_a\leq x\leq x_b)\rangle_t=\int_{x_a}^{x_b}P_t(x)\Intd x} is the probability of finding a particle between \eq{x_a} and \eq{x_b} at time \eq{t}. At long-times, this is asymptotically equal to \eq{\sim ({1}/{\mathcal{Z}_t})\int_{x_a}^{x_b}e^{-V(x)/k_BT}\Intd x,} which decays in time as \eq{t^{-1/2}} via \eq{\ZZ_t}, reflecting the diffusion of particles out to infinity.} 
Note that the behavior of non-integrable observables such as the mean-squared displacement (MSD) \eq{\langle x^2\rangle_t\sim 2 D t}, is similar to Brownian motion, since such observables are determined by the Gaussian tails of the probability density. 

\textit{Time-averaged observables.} 
When the standard Boltzmann-Gibbs framework does not apply, time-averages are random even in the long-time limit. Here we will provide their distributional limiting behavior, inspired by infinite-ergodic theory. 
We start with the ensemble-time-average, defined as follows:  \eq{\langle\overline{\OO[x(t)]}\rangle=\langle(1/t)\int_0^t\OO[x(t')]\Intd t'\rangle}. {To measure this experimentally, one should obtain the time-average from single particle trajectories, and further average over many paths.} Using Eq.~\eqref{EA}, for an observable integrable with respect to the non-normalizable Boltzmann state, after replacing the order of the ensemble- and the time- averaging procedures and neglecting short-time effects, we obtain  \EQ{\left\langle\overline{\OO[x(t)]}\right\rangle\sim\frac{1}{t}\int_0^t\langle\OO(x)\rangle_{t'}\Intd t'=2\langle\OO(x)\rangle_{t}.}{TA} 
Thus the value obtained for the average quantity depends on the method of measurement, and even though at long times the ratio \eq{\langle\overline{\OO[x(t)]}\rangle/\langle \mathcal{O
}(x)\rangle_t}  
 approaches a constant, it is twice the value expected from ordinary statistical mechanics.
This doubling effect is the result of the integration over the time-dependent {\eq{1/\ZZ_t\propto t^{-1/2}}}.  In Fig. \ref{fig2}{, we show that as \eq{t\rightarrow\infty}, \eq{\langle\overline{V_{LJ}[x(t)]}\rangle/\langle V_{LJ}(x)\rangle_t\rightarrow2}}, where the time-average was obtained from simulations of particles in the LJ potential [28], and the ensemble-average was calculated using Eq. \eqref{EA}.  
\begin{figure}[t]
\centering 
\includegraphics[width=1.0\linewidth]{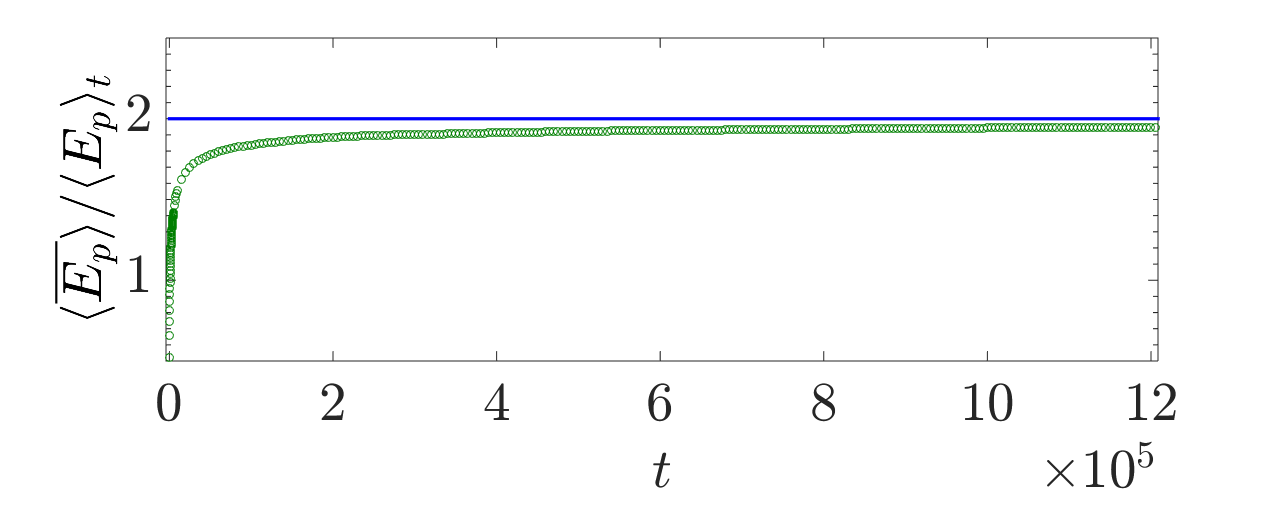}
\caption{{\footnotesize{(Color online). The ratio between the ensemble-time-averaged potential energy \eq{\langle\overline{E_p}\rangle=\langle\overline{V_{LJ}[x(t)]}\rangle}, obtained from simulations [28] with the LJ-type potential, Eq. \eqref{Potentials}, and the ensemble-mean \eq{\langle{E_p}\rangle_t=\langle{V_{LJ}(x)}\rangle_t}, obtained from Eqs. (\ref{Potentials},\ref{ID},\ref{EA}) (green, circles). The parameters of the potential and the temperature are the same as in Fig. \ref{fig1}. The asymptotic limit  (blue line) is given by Eq. \eqref{TA}. The same effect is found also for other integrable observables, e.g., \eq{\Theta(\cdot)}.
}}}
\label{fig2}
\end{figure}  
Note that, for a non-integrable observable such as the MSD (see also \cite{akimoto2008generalized,akimoto2015distributional}), we use the uniform approximation Eq. \eqref{UniformPDF}, which yields  \eq{\langle \overline{x(t)^2}\rangle\approx Dt}, which is again similar to Brownian motion \cite{andreanov2012time}. 

\begin{figure}[t]
\centering
\includegraphics[width=1.0\linewidth]{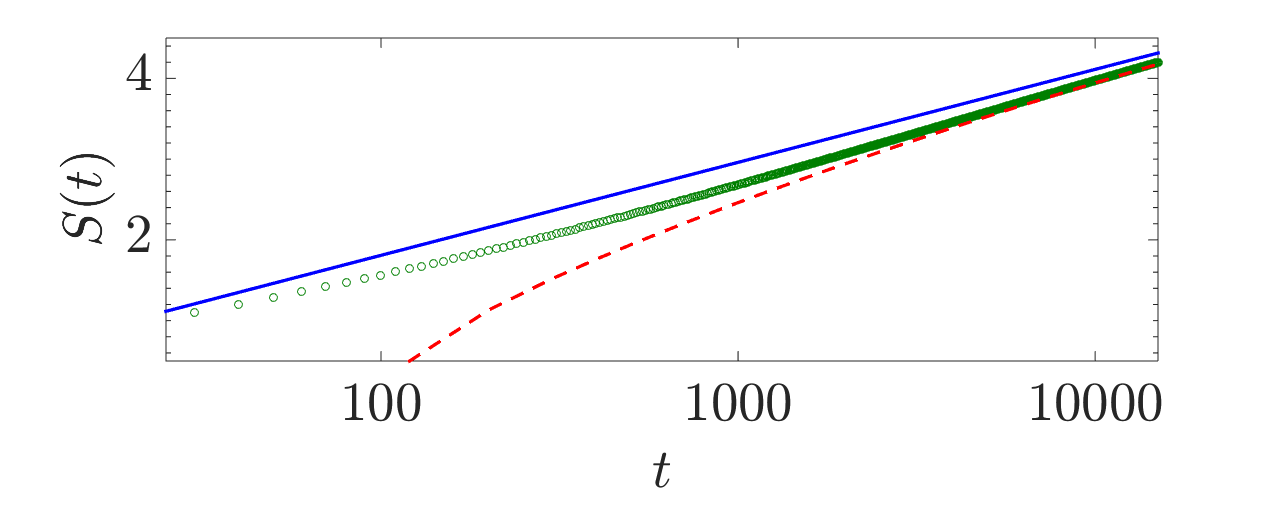}
\caption{{\footnotesize{(Color online) The Gibbs entropy, obtained from simulation results, with a LJ potential as in Fig. \ref{fig1}, versus time (green circles). The theoretical curve (red dashed line) corresponds to Eq. \eqref{Entropy}, where the averaged potential energy is obtained from Eq. \eqref{EA}, and the MSD is obtained from the uniform approximation, Eq. \eqref{UniformPDF}. At the long \eq{t} limit, $S(t)\sim (k_B /2)\ln(\pi e D t)$ (blue line)
.}}}
\label{fig4}
\end{figure}

{\em Entropy extremum.} 
To create a framework for the above non-normalizable Boltzmann statistics, we now formulate a maximum entropy
principle, in the spirit of thermodynamics, with the
additional constraint that  \eq{\langle x^2\rangle_t\sim 2Dt}, at the long-time limit. 
The functional to be maximized includes four terms
\begin{align}
&S[P_t(x)]=-k_B\int_0^\infty P_t(x)\ln(P_t(x))\Intd x\nonumber\\ 
&-k_B\rho\left[\int_0^\infty P_t(x)\Intd x-1\right]-k_B\beta\left[\int_0^\infty V(x)P_t(x)\Intd x-\langle E_p\rangle\right]\nonumber\\ 
&-k_B\zeta\left[\int_0^\infty x^2P_t(x)\Intd x-2Dt\right].
\label{}
\end{align}
The first term is the Gibbs entropy, and $\rho$,  $\beta$ and
$\zeta$ are Lagrange multipliers associated with the  normalization, average energy and MSD, respectively. {The constraint on the MSD means that the central-limit theorem is dominant at large distances from the surface, where the potential is vanishing.}  
Taking the functional derivative, we obtain the uniform approximation, Eq. \eqref{UniformPDF} [26].  
The entropy, obtained from the uniform approximation, Eq. \eqref{UniformPDF},
reads 
\EQ{S(t)=k_B{\ln{(\pi D t)}}/2+\langle E_p\rangle/T+k_B\langle x^2\rangle_t/4Dt.}{Entropy}  
This relation is demonstrated by comparing simulation results versus theory in Fig. \ref{fig4}. Importantly, it shows that, for a fixed observation time; {$1/T = (\partial S/\partial U)_t$}, in agreement with {standard} thermodynamics. 
As expected, in the very long time limit  entropy increases logarithmically in time, since the number of states is infinite and
\eq{\mathcal{Z}_t} is time dependent. 

{\textit{Virial theorem.}} 
As seen by the maximum entropy principle,  the toolbox of thermodynamics can be extended to non-normalizable Boltzmann-Gibbs statistics. To further demonstrate this, we studied the virial theorem, which addresses the mean of the observable \eq{xF(x)} (\eq{F(x)=-V'(x)}). Strongly binding potentials, treated with standard thermodynamics,  yield \eq{\langle xF(x)\rangle=-k_BT}. In our case, since it is an integrable observable, for asymptotically flat potentials  {\EQ{\langle x F(x)\rangle_t\approx \frac{2B_2}{\mathcal{Z}_t}k_BT,\Hquad B_2=\frac{1}{2}\int_0^\infty\left[1-e^{-V(x)/k_BT}\right]\Intd x}{Virial}}at long times, where \eq{B_2} is the second virial coefficient \cite{Chandler}. {This result is obtained via 
\eq{\langle x F(x) \rangle \approx(k_BT/\mathcal{Z}_t)\intֹ_0^\infty x  \HHquad \partial_x[\exp(- V(x)/k_BT)-1]\Intd x}, noting that 
\eq{[\exp(-V(x)/k_BT)-1]\rightarrow0} when \eq{x\rightarrow\infty}.}  

\begin{figure}[t]
\centering
\includegraphics[width=1.0\linewidth]{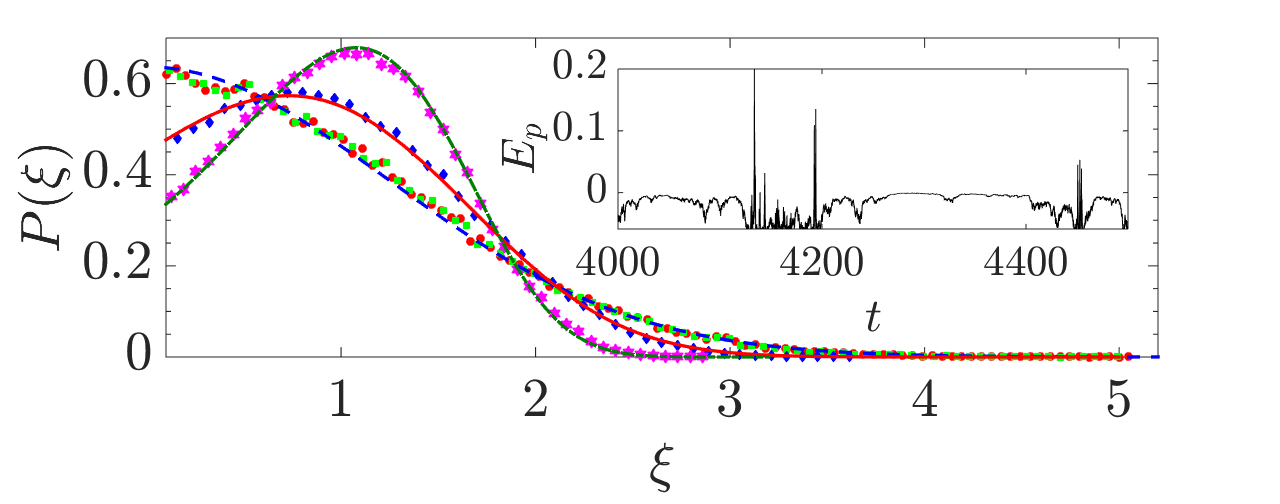}
\caption{{\footnotesize{(Color online). The probability density \eq{P(\xi)}, where \eq{\xi = \overline{{\cal O}}/
\langle \overline{{\cal O}} \rangle}, 
approaches in the long time limit to the Mittag-Leffler distribution, Eq. \eqref{MLDistribution}. The theory is \eq{\mathcal{M}_\alpha(\xi)}, with ${\alpha=0.5}$, which is equivalent to half a Gaussian (dashed blue line), ${\alpha=0.64}$ (red line), and ${\alpha=0.75}$ (green dash-dot line). 
The observables obtained from the LJ potential are: potential energy, ${\xi=\overline{V}/\langle\overline{V}\rangle}$ (green squares), and the indicator function (defined as unity for \eq{x\in[0,5]} [28]) ${ \xi=\overline{\Theta} /\langle\overline{\Theta}\rangle}$  (red circles). Simulation results for the indicator function, with a log potential and \eq{\alpha=0.64,0.75} (\eq{x\in[0,3.2]} and \eq{[0,8]} respectively), appear in (blue diamonds) and (magenta stars) respectively [28]. 
\textit{Inset:} A time series of the potential energy of a single particle in ${V_{LJ}(x)}$, demonstrating short periods at which the particle samples the region of non-negligible potential, with long sojourn times in-between. }}}
\label{fig3}
\end{figure} 

Since the non-normalizable Boltzmann state can be used to obtain information on 
physically important observables like energy, entropy, the virial theorem and occupation times, we now turn to 
the discussion of the distribution of the time-averages, which will replace standard
ergodicity. 

\textit{Distribution of time-averages.}  
Here we study the {distribution} of the random time-average \eq{\overline{\OO[x(t)]}}, for an integrable observable. Due to the attractive potential, the  unidimensional processes we are studying are recurrent, however they are characterized by two distinct time scales. Consider first, for simplicity, the case of the observable \eq{\Theta(x_a\leq x(t)\leq x_b)}. The time that the particle spends inside the finite region \eq{[x_a,x_b]} at each visit has a finite typical value \eq{\langle t_{in}\rangle}. 
However once the particle experiences a large thermal fluctuation, taking it to large \eq{x}, the probability distribution of its first-return time to this region is fat-tailed; \eq{f(t_{ret})\sim t_{ret}^{-1-\alpha}} \cite{bray2000random,lutz2004power}. If the potential is asymptoticly flat \eq{\alpha=1/2} and \eq{\langle t_{ret}\rangle} diverges. This blow-up of the typical return time is the physical reason why standard ergodic theory is not applicable here, since for a time-averaged observable to converge to its ensemble-mean, the measurement time must be long compared to the microscopical scale. 

Therefore the random occupation time \eq{\tau={\int_0^t{\Theta(x_a\leq x(t')\leq x_b)}\Intd t'}}, which is a functional of the particle's trajectory \eq{x(t')}, is proportional to \eq{\langle t_{in}\rangle n}, where \eq{n} is the random number of \eq{x_b} crossings into the region \eq{[x_a,x_b]}, until time \eq{t}.  
Since it is proportional to \eq{n}, we have
\eq{\tau/\langle \tau\rangle\rightarrow n/\langle n\rangle=\xi}. The distribution of this ratio, \eq{\mathscr{M}_\alpha(\xi)}, is the well-known 
 Mittag-Leffler distribution:  
\begin{equation}
 \mathscr{M}_\alpha(\xi)= \frac{\Gamma^{1/\alpha}(1+\alpha)}{\alpha \xi^{1+1/\alpha}}L_\alpha\left[\frac{\Gamma^{1/\alpha}(1+\alpha)}{\alpha \xi^{1/\alpha}}\right],
 \label{MLDistribution} 
\end{equation} 
obtained from renewal theory for a process with power-law sojourn times  \cite{feller1971william,aaronson1997introduction}. Here \eq{\Gamma(\cdot)} is the Gamma function and \eq{L_\alpha[\cdot]} is the one-sided L\'evy density \footnote{\eq{L_\alpha[x]}, where \eq{x>0}, is defined as the inverse-Laplace transform of  \eq{\exp(- u^\alpha)}, from \eq{u\rightarrow x}}.   
The logic behind this result is that the crossing times of \eq{x_b} are not correlated, and that the time spent in \eq{(0,x_a)} is statistically much shorter than the time in \eq{x>x_b}. 
The above arguments, according to the Aaronson-Darling-Kac theorem \cite{aaronson1997introduction}, apply
to any observable integrable with respect to an infinite measure of a system  \cite{Aaronson2005ergodic}, and by extension also to our non-normalizable Boltzmann state, so {the distribution \eq{P(\xi)}, where \eq{\xi=\overline{\OO[x(t)]}/\langle\overline{\OO[x(t)]}\rangle}, is equal to \eq{\mathcal{M}_\alpha(\xi)}, Eq. \eqref{MLDistribution}}, and \eq{\alpha} is related to the first return-time statistics. 
{The probability densities of the indicator function and the time-averaged potential energy,  in the presence of the LJ potential, are shown in Fig. \eqref{fig3}, along with the Mittag-Leffler distribution. A sample of the time series of \eq{E_p} is presented in the inset of  Fig. \ref{fig3}, showing rare ``renewal'' events, at which the value of the energy is non-negligible (namely its absolute value is above an arbitrary threshold \eq{\epsilon\ll1}). This provides the intuitive explanation for why this observable can be treated in analogy with the occupation-time.}  

Our theory  extends also to the case of logarithmic potentials, e.g.  \cite{campisi2012logarithmic,hirschberg2012diffusion,wawrzkiewicz2015application,kim2015distribution}. This case provides an additional important insight into the  distribution of the return times, and the connection between infinite-ergodic theory and Boltzmann-Gibbs statistics. As shown in \cite{dechant2011solution}, for potentials where \eq{V(x)\sim V_0\log(x/l_0)} at large~\eq{x};  \eq{\lim_{t\rightarrow\infty}  \mathcal{Z}_tP_t(x)=e^{-V(x)/k_BT}}, where \eq{\ZZ_t=(\Gamma(1-\alpha)l_0/2) (4Dt/l_0^2)^{1-\alpha}}, and  \eq{\alpha=1/2+V_0/(2k_BT)}. 
Here, the asymptotic thermal state is non-normalizable at sufficiently high temperatures, when~$0<V_0/k_B T<1$, namely  \eq{1/2<\alpha<1}. To derive the non-normalized Boltzmann state for logarithmic potentials from entropy maximization, one should use the MSD obtained in }\cite{dechant2011solution}; \eq{\langle x^2\rangle_t\sim 4D(1-\alpha)t}, which also yields the time-dependent entropy-energy relation in this case. This will be published elsewhere. For integrable observables, using a similar derivation as for asymptotically flat potentials, we now find that Eq. \eqref{EA} is still valid, and \eq{\langle\overline{O[x(t)]}\rangle/\langle O(x)\rangle_t\rightarrow1/\alpha,} when \eq{t\rightarrow\infty}.  
When \eq{\alpha\rightarrow1}, the Boltzmann factor becomes normalizable, and infinite-ergodic theory reduces to the standard ergodic hypothesis, where \eq{\ZZ_t} is replaced by the normalizing partition function \eq{Z}. 
 In the SM we test our theory, showing its predictions perfectly match numerical simulations. 
With a logarithmic potential, \eq{\alpha} in Eq. \eqref{MLDistribution} can be tuned in the range \eq{1/2\leq \alpha<1}, for example by adjusting the temperature,  to obtain a smooth transition of \eq{\mathscr{M}_\alpha(\xi)} from half a Gaussian corresponding to \eq{T\rightarrow\infty}, to ergodic behavior at \eq{T\rightarrow V_0} (as mentioned, beyond this value, the Boltzmann factor is normalizable). This is demonstrated in Fig. \ref{fig3}.
Note that with a logarithmic potential, in the low temperature regime, the properly time-scaled power-law tails of the normalized density \eq{P_t(x)} are described by an infinite-covariant density  \cite{kessler2010infinite,dechant2011solution}, however this does not lead to infinite-ergodic theory.

{\textit{Generality.}} 
Infinite-invariant measures can be found in any dimension, as is easy to verify for free  \eq{d}-dimensional Brownian motion. This property is preserved also in the presence of an asymptotically weak, or logarithmic potential, similarly to the unidimensional case. We will discuss this at length in a future publication (see SM). Clearly, infinite-ergodic theory, and the Mittag-Leffler distribution of time-averaged observables, require the process to be recurrent. When the first-passage time distribution follows a power-law with a diverging mean, infinite-ergodic theory applies. 
In 1-dimension, when the potential field is unstable, e.g. \eq{V(x) \propto x^3},  
the process is non-recurrent, and a different theory emerges  \cite{martin2018diffusion,ornigottibrownian}. 

Extending infinite-ergodic theory to the underdamped generalized Langevin equation, which is non-Markovian, is a worthy goal. This process is also recurrent, and the fluctuation-dissipation theorem holds, so we expect the main properties of our theory to remain. Furthermore, the single-particle analysis used in this letter reflects current day single-molecule experiments; however our work can be elevated to a many-body theory, {which could be related to the process described in  \cite{\iffalse Chafai2018Simulating,\fi mithun2018weakly}}. There, the potential energy of the single particle is replaced by the many-body potential, but the non-normalizable Boltzmann factor remains. 
Infinite-ergodic theory can also be extended to many other systems, for example Langevin equations with a multiplicative noise \cite{NavaFuture}, and diffusion on fractals
. 
\\

The support of Israel Science Foundation grant number $1898/17$ is acknowledged.

\bibliographystyle{aipnum4-1}
\bibliography{./bibliography2} 

\begin{widetext} 
\newpage 
\renewcommand{\baselinestretch}{1.5} 
\setcounter{figure}{0}
\renewcommand{\figurename}{SM}

\onecolumngrid

\begin{center}\textbf{{Supplemental material for:\\ From Non-normalizable Boltzmann-Gibbs statistics to infinite-ergodic theory}}\end{center}
%
\section{A. 3-dimensional Langevin motion near a wall, with a potential \eq{V(x)}}

 The 3-dimensional motion of a particle in a  thermal bath, floating above a flat surface (blue curve). Here, \eq{x>0} is the distance between the particle and the surface.    
The surface-single molecule potential 
\eq{V(x)} is provided in Eq. (2) in the main text, and shown in the inset of Fig. 1 there. The motion parallel to the \eq{y-z} plane is purely diffusive (in the figure we modeled it as a simple symmetric coin-toss random walk, independently along each axis). The inset shows the trajectory \eq{x(t)} (black line). To simulated the trajectory for the \eq{\hat{x}} axis, we used  an Euler-Mayurama integration of Eq. (1) in the main (where  \eq{(\gamma,k_BT)=(1,0.0469)}), with the Lennard-Jones potential, Eq. (3) (with \eq{V_0=1000}, \eq{a=2} , \eq{b=1}), and with time increments of \eq{\Delta t=0.01} (\eq{100000} time steps). 

\begin{figure*}[h]
\centering
\includegraphics[width=0.7\linewidth]{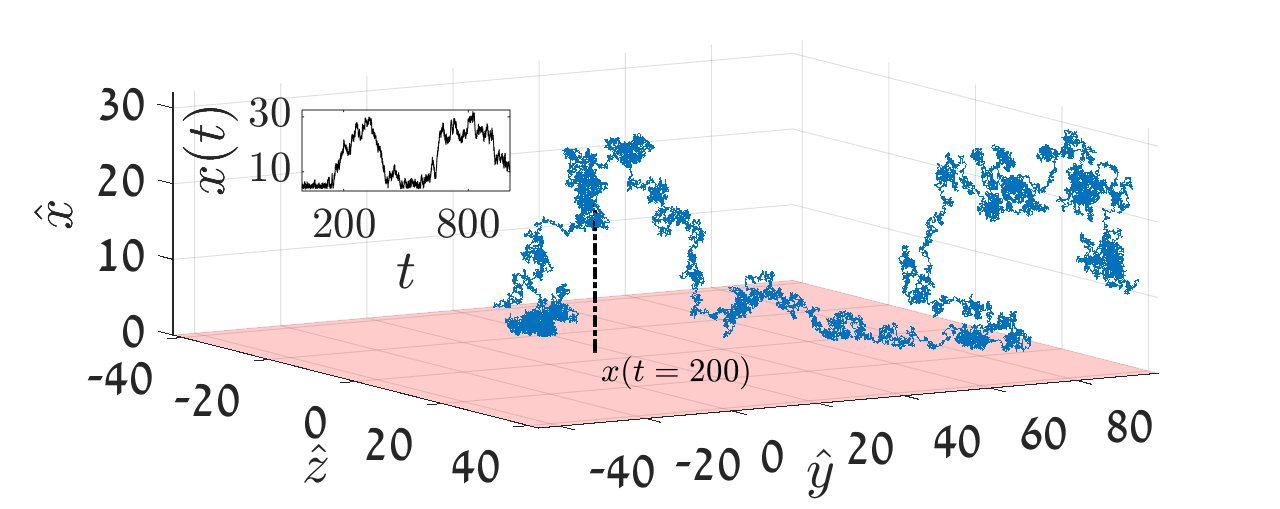}
\caption{{\footnotesize{(Color online) An example of the Langevin trajectory of a particle diffusing freely in the \eq{y-z} plane, and following Eq. (1) in the main text, along the \eq{\hat{x}}-axis (blue line), with the Lennard-Jones potential, Eq. (3) (in the main text). The inset shows the trajectory \eq{x(t)} (black line).}}}
\label{SM0}
\end{figure*} 
\section{B. Derivation of the non-normalizable Boltzmann state, using a scaling ansatz} 
We now derive the non-normalizable Boltzmann state from the Fokker-Planck Eq. (2) in the main text, in the case of asymptotically flat potentials, using a scaling ansatz. 
The Fokker-Planck equation is:  \EQ{\frac{\partial P_t(x)}{\partial t}=\frac{D}{k_BT}\frac{\partial }{\partial x}V'(x)P_t(x) +D\frac{\partial^2 }{\partial x^2}P_t(x).}{Kramers} 
Since \eq{V'(x) = 0} for large \eq{x} the process is diffusive,  and the tails of the probability density fall-off as \eq{t^{-1/2}\exp(-x^2/4Dt)}, we make the scaling ansatz: 
\EQ{P_t(x)\propto t^{-1/2}e^{-x^2/(4D t)}\mathcal{I}(x).}{ScalingAnsatz}  
Substituting this in the left-hand side of Eq. \eqref{Kramers}, we obtain: 
\EQ{-(1/2) t^{-3/2}e^{-x^2/(4D t)}\mathcal{I}(x)+t^{-5/2}\frac{x^2}{4D}e^{-x^2/(4Dt)}\mathcal{I}(x).}{LHS1} 
On the right-hand side, we have 
\begin{align}
&\frac{D}{k_BT}\left[-V'(x)t^{-3/2}\frac{x}{2D}e^{-x^2/(4Dt)}\mathcal{I}(x)+V''(x)t^{-1/2}e^{-x^2/(4D t)}\mathcal{I}(x)+V'(x)t^{-1/2}e^{-x^2/(4Dt)}\frac{\partial}{\partial x}\mathcal{I}(x)\right]+\nonumber\\ 
& D\left[t^{-5/2}\left(\frac{x}{2D}\right)^2e^{-x^2/(4Dt)}\mathcal{I}(x)-t^{-3/2}\frac{x}{D}e^{-x^2/(4D t)}\frac{\partial}{\partial x}\mathcal{I}(x)-t^{-3/2}\frac{1}{2D}e^{-x^2/(4Dt)}\mathcal{I}(x)+t^{-1/2}e^{-x^2/(4D 
t)}\frac{\partial^2}{\partial x^2}\mathcal{I}(x)\right].
\label{RHS1}  
\end{align} 
Comparing terms of leading order in time, namely proportional to \eq{t^{-1/2}}: 
	\EQ{-V''(x)\mathcal{I}(x)-V'(x)\frac{\partial}{\partial x}\mathcal{I}(x)\sim k_BT\frac{\partial^2}{\partial x^2}\mathcal{I}(x).}{tAlpha} 
This equation has two solutions: 
\EQ{\mathcal{I}_1(x)\propto e^{-V(x)/k_BT},\qquad\mbox{and}\qquad \mathcal{I}_2(x)\propto e^{-V(x)/k_BT}\int_{1}^x e^{V(x')/k_BT}\Intd x'.}{Solutions}
 Here, the first solution, \eq{\mathcal{I}_1(x)}, is the non-normaliable Boltzmann state, to which the probability density of the particles converges at \eq{t\rightarrow\infty}, via Eq. (5) in the main text. The second solution, \eq{\mathcal{I}_2(x)}, is not physical, since it cannot be matched with the Gaussian fall-off at the tails of \eq{P_t(x)}, Eq. \eqref{ScalingAnsatz}. 

\section{C. Entropy maximization
}


Here we add a few technical details on the derivation of the maximum entropy  principle, presented in the main text. 
From the functional derivative of Eq. (8) in the main text, we get
\begin{equation}
-\ln[P_t(x) ]- 1 - \rho - \beta V(x) - \xi x^2 =0
\end{equation}
Hence
$$ P_t(x) = \exp[ - \beta V(x) - \xi x^2]/\mathcal{Z}_t,$$ 
with $\mathcal{Z}_t = \exp( 1 + \rho)$.
\eq{\beta=1/k_BT} as usual, and  from the normalization constrain
$$ \mathcal{Z}_t =\int_0 ^\infty  \exp[ - \xi x^2 - \beta V(x) ] {\rm d} x. $$
The mean-squared displacement is
\begin{equation}
\langle x^2 \rangle_t= {\int_0 ^\infty x^2  \exp[ - \xi x^2 - \beta V(x)] {\rm d} x \over \mathcal{Z}_t}
\end{equation}
Given that $\langle x^2 \rangle_t
\sim 2 D t$ in the long time limit, we find
$\xi\sim 1/(4 D t)$. In addition,  $\rho=\ln[\mathcal{Z}_t/e]$, which  in the long time limit yields 
 $\mathcal{Z}_t \sim \sqrt{\pi D t}$, since the contribution from the regime of \eq{x} where the potential is non-negligible, is exponentially small. 
\section{D. Non-Normalizable Boltzmann state for logarithmic potentials} 

In Fig. SM2 we present the non-normalized Boltzmann state \eq{\exp(-V_{log}(x)/k_BT)}, where the logarithmic potential is \eq{V_{log}(x)=V_0\log(1+x^2)/2}. \eq{\mathcal{Z}_tP_t(x)} is also shown. \eq{P_t(x)} is the probability density obtained from simulation results, performed using an Euler-Mayurama integration of the Langevin Eq. (1) in the main text, with \eq{\gamma=1}, \eq{V_0=0.14}, 
\eq{k_BT=0.5} and time-steps of duration \eq{\Delta t=0.01}. Here \eq{t=59000}, \eq{\mathcal{Z}_t=\Gamma(1-\alpha)(4D t)^{1-\alpha}/2}, and \eq{\alpha=1/2+V_0/(2k_BT)}. The number of particles is \eq{10^6}. The simulation result agrees with the non-normalized Boltzmann state, without any fitting. The inset shows the potential. In Fig. SM3, we show the ratio \eq{\langle\overline{\Theta(\cdot)}\rangle/\langle{\Theta(\cdot)}\rangle_t} for particles in \eq{V_{log}(x)}, versus time, for various \eq{\alpha}s, and in various regimes of \eq{x}, confirming the asymptotic approach of this ratio to \eq{1/\alpha}, as explained in the main text. 

\begin{figure*}[h]
\centering
\includegraphics[width=0.7\linewidth]{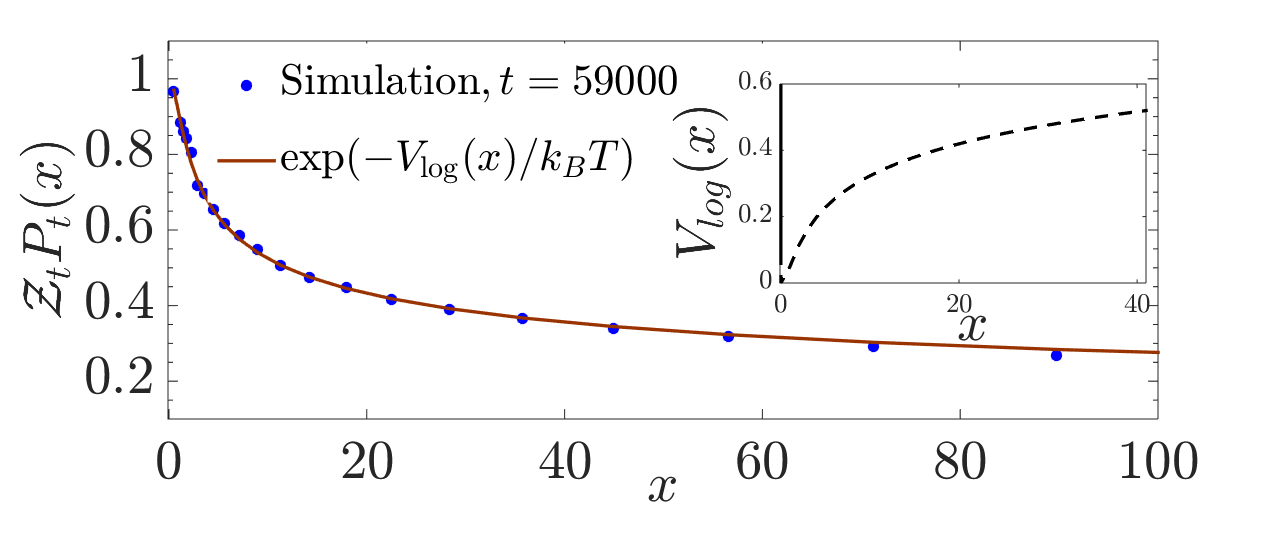}
\caption{{\footnotesize{(Color online) The non-normalizable Boltzmann state for \eq{V_{log}(x)} (brown, solid line). The  simulation result at \eq{t=59000} is presented in (blue, circles).  The presented solution is not normalized, since \eq{\mathcal{Z}_t P_t(x)\sim x^{-V_0/k_BT}}
which is non-integrable for large \eq{T}. The inset shows the potential (dashed line).}}}
\label{SM2}
\end{figure*}  

\begin{figure*}[h]
\centering
\includegraphics[width=0.7\linewidth]{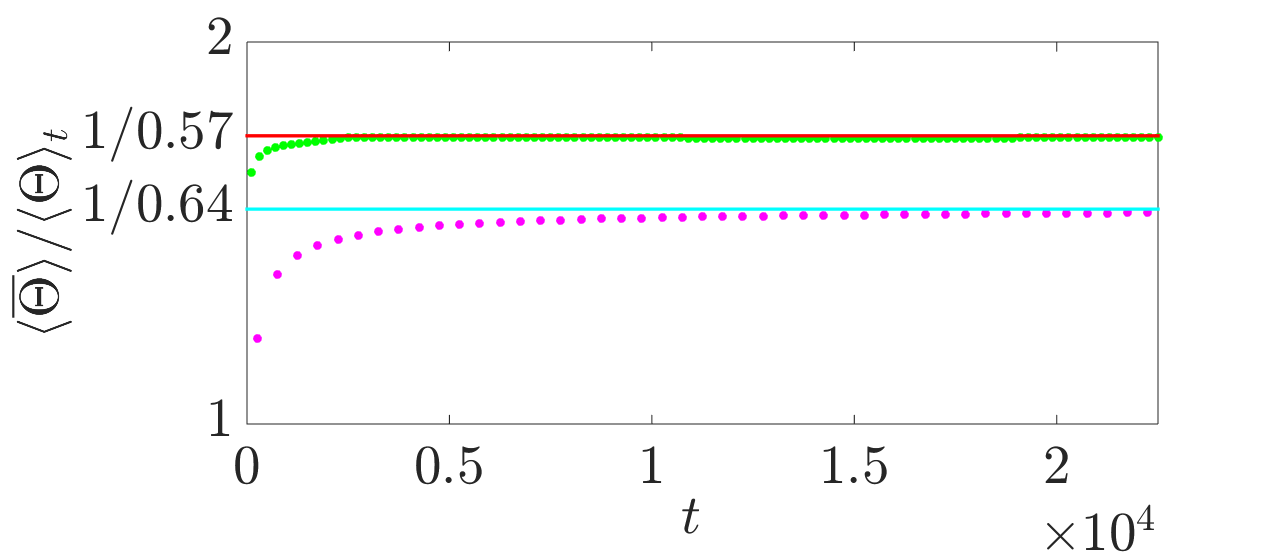}
\caption{{\footnotesize{(Color online) The ratio \eq{\langle\overline{\Theta(\cdot)}\rangle/\langle{\Theta(\cdot)}\rangle_t} for particles in \eq{V_{log}(x)}, versus time. For \eq{\alpha=0.64} and \eq{\alpha=0.57}: simulations appear in (magenta, circles) and (light-green, circles), respectively, and the asymptotic limit \eq{1/\alpha}, approached as \eq{t\rightarrow\infty}, in  (cyan line) and (red line), respectively. For \eq{\alpha=0.64}, simulations were performed with \eq{(V_0,k_BT)=(0.14,0.5)} and \eq{x\in[0,3.2]}. For \eq{\alpha=0.57}, \eq{(V_0,k_BT)=(0.8,5.71)} and \eq{x\in[0,5]}.}}}
\label{SM4}
\end{figure*}  

\newpage
\section{E. Simulation in $d$ dimensions}

\begin{figure*}[h]
\includegraphics[width=0.55\linewidth]{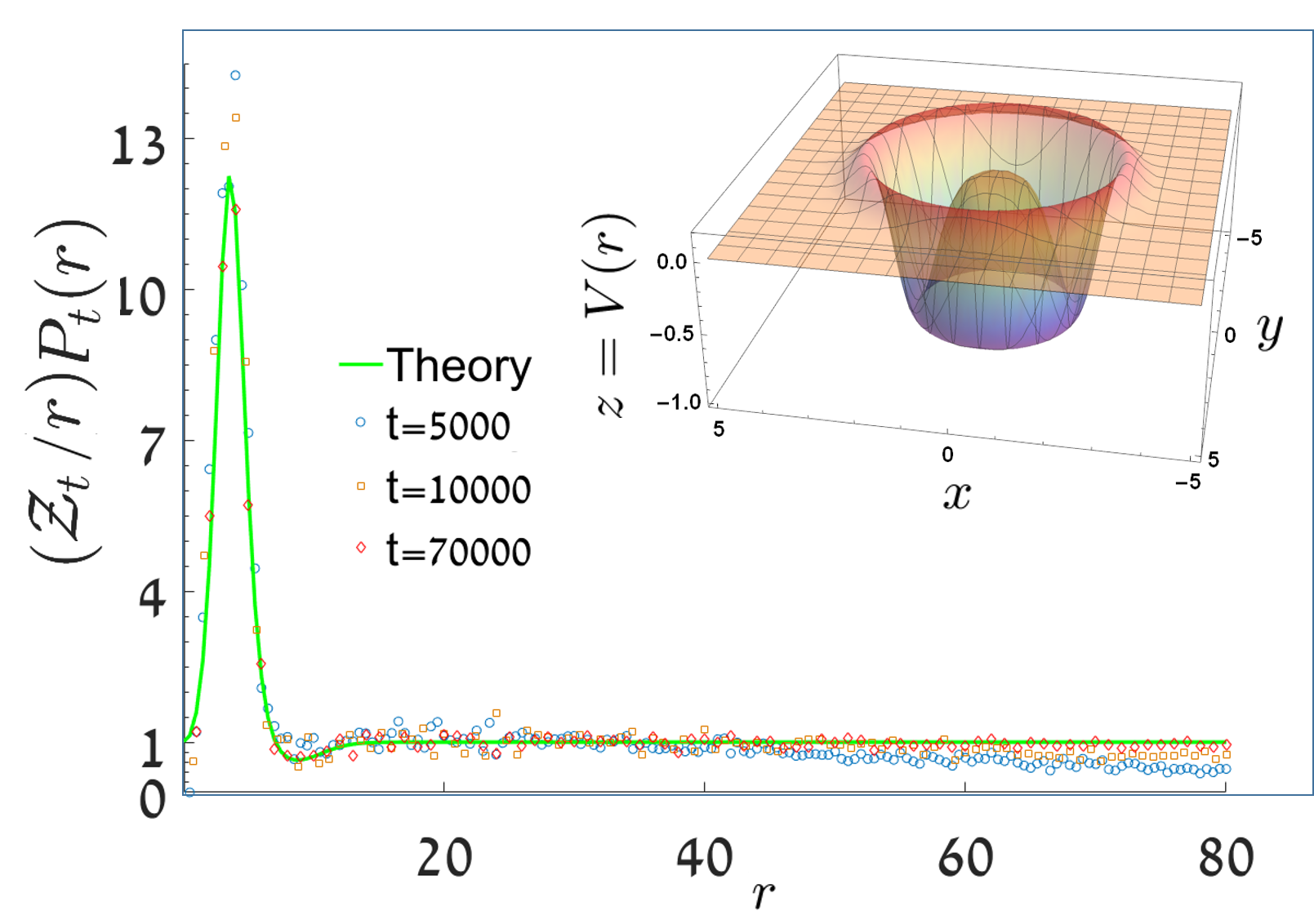}
\caption{{\footnotesize{(Color online) Simulation results for a two-dimensional Langevin process, in the \eq{x-y} plane, with \eq{V(r)=\left((-25r^2 + r^4/2)/125\right)\exp(-r^2/20)}, versus theory. We chose \eq{\gamma=1}, and \eq{k_BT=D=0.4}. In colored symbols:  \eq{(\mathcal{Z}_t/r)P_t(r)}, where \eq{r=\sqrt{x^2+y^2}}, \eq{\mathcal{Z}_t=2D t} and \eq{P_t(r)} is the radial distribution, obtained from the simulation at times \eq{t= 5000, 10000,70000}, appear in (blue circles, brown squares and red diamonds), respectively. The non-normalizable Boltzmann factor \eq{\exp(-V(r)/k_BT)}, appears in (green line). The inset shows the potential (the rainbow color-map illustrates the height differences, blue means potential well, red means hill).}}}
\label{SM3}
\end{figure*}  


In Fig. SM4, we demonstrate the approach of \eq{(2Dt/r)P_t(r)}, where \eq{P_t(r)} is the normalized probability density of the radial distance \eq{r} of the particles from the origin, for a two-dimensional process with an isotropic potential field \eq{V(r)=\left((-25r^2 + r^4/2)/125\right)\exp(-r^2/20)}, at increasing times, to the non-normalizable Boltzmann state, \eq{\exp(-V(r)/k_BT)}. Simulation results are presented in (colored symbols), and the theory in a (green line).  As is well-known to mathematicians, in the long-time limit, free Brownian motion in $d$ dimensions is described by an infinite-invariant density. However, extending this to the case of Langevin processes in the presence of a force-field, as seen in Fig. SM4  is of higher interest physically. This will be discussed in detail in a future publication.  

\section{F. Some simulation parameters, for the figures in the main text}

All the simulations in this letter were performed as standard Euler-Mayurama integration of Eq. (1), in the main text,  
with \eq{\gamma=1} and a time-step duration of \eq{\Delta t=0.01} (in dimensionless units), over \eq{10^5-10^6} particles. For the LJ (logarithmic) potential;  \eq{t=10^5\Delta t-10^8\Delta t} (\eq{t=6\times10^6\Delta t}). 
Here are the particular details regarding Fig. 4 in the main text:  
The simulation details for ${\alpha=0.5,0.64,0.75}$ (red circles,blue diamonds, magenta stars) respectively, are: for ${\alpha=0.5}$, similar to Fig. 1 in the main text (the occupation-time was obtained in the range \eq{x\in[0,5]}), at time \eq{t=10^8\Delta t}, and for ${\alpha=0.64}$ we used ${V_{log}(x)}$ (defined in Sec. C. of this supplemental material), with ${(V_0,k_BT)=(0.14,0.5)}$, ${x\in[0,3.2]}$. For ${\alpha=0.75}$, we used the logarithmic potential, with  
${(V_0,k_BT)=(0.625,1.25)}$, ${x\in[0,8]}$. 
\end{widetext}

\end{document}